\documentclass[12pt,final,epsfig]{elsart}

\usepackage{epsf} 
\usepackage{natbib} 
\usepackage{graphicx}
\usepackage{amsmath} 
\usepackage{amsxtra} 
\usepackage{amssymb}
\usepackage{amsfonts} 
\usepackage{amstext} 
\usepackage{amsgen} 
\usepackage{amsbsy} 
\usepackage{amsopn} 
\usepackage{amscd}

\begin{document}

\begin{frontmatter}

\title{An Advanced Analysis Technique for Transient Searches in Wide-Field Gamma-Ray Observatories}
\author[SCIPP,MIT]{Miguel F. Morales},
\author [SCIPP]{David A. Williams} and
\author [SCIPP,UMd]{Tyce DeYoung}

\address [SCIPP]{Santa Cruz Institute for Particle Physics, University of California Santa Cruz, CA  95064}

\address [MIT]{Center for Space Research, Massachusetts Institute of  Technology, MA  02139}

\address [UMd]{Department of Physics, University of Maryland, MD 20742}

\begin{abstract} Wide-field gamma-ray telescopes typically have highly variable event-by-event resolution which leads to a number of unique and challenging analysis requirements --- particularly when conducting transient searches over multiple time scales.  By generalizing the ideas of the Gaussian weighting analysis to point-spread functions of  arbitrary shape and the regime of Poisson statistics, an efficient analysis which uses the event-by-event resolution is developed with a  sensitivity similar to that of a well-implemented maximum likelihood analysis. In this development, the effect of a number of different  approximations on the sensitivity and speed of the final analysis can be easily determined and tuned to the particular application.  The analysis method is particularly well suited to transient detection in wide field-of-view gamma-ray observatories, and is currently used for the 40 s -- 3 hour transient search in the Milagro observatory.

\end{abstract}


\end{frontmatter}

\section{Introduction}

The observed point spread function (PSF) in many gamma-ray observatories can vary tremendously from one event to the next, depending on the characteristics of the individual event.  In Milagro  the observed PSF width varies by an order of magnitude from the best  events to the worst \citep{MyThesis} and the GLAST PSF is expected to  vary by more than two orders of magnitude.  The PSFs typically depend on the characteristics of the individual events, such as the number of  interactions within the detector and the energy of the initiating particle, and these characteristics can be used to determine the PSF of the particular event.  This strong dependence of the PSF on event  parameters is a feature of many gamma-ray observatories, and significantly complicates the data reduction.  

Point source search techniques used in high energy astrophysics are reviewed in \citet{CygnusTeq}, with optimal bin and maximum likelihood  analysis being the most common techniques. Maximum likelihood analyses  are the traditional choice for variable PSF instruments, and while they can be computationally demanding they do make effective use of the event characteristics and variable PSF information.  Optimal binned analyses, on the other hand, are very fast but are less sensitive due to their inability to effectively use the individual event characteristics.  The analysis method presented in this paper was developed to provide a compromise between optimal bin and maximum  likelihood analyses, and is particularly useful for analyses in variable PSF instruments where both sensitivity and speed are important.

In an optimal binned analysis the sky is divided into equal area bins with size chosen to maximize the signal-to-noise ratio for the detector's average PSF.  The number of events observed in each bin is simply counted, and the Poisson probability determines the significance of observed signals.  In essence, optimal binned analyses  ignore the information provided by the characteristics of individual events, and instead treat all events as if they were drawn from the average PSF distribution. While statistically correct, ignoring the event quality information reduces the sensitivity of the analysis (see  Section \ref{WATSensitivity}). The advantage of an optimal bin analysis is its simplicity and speed. 

Maximum likelihood techniques can use the information from the individual event characteristics and are the most powerful methods for analyzing wide-field gamma-ray observations.  However, most implementations are computationally demanding because they require fitting parameters of the likelihood model, and this fit usually involves an iterative fitting algorithm such as MINUIT \citep{MINUIT}. This can make maximum likelihood unsuitable for applications where computational efficiency is an important constraint.

A third analysis method is the Gaussian weighting technique which has been used by the Fly's Eye and JANZOS experiments as a compromise between optimal bin and maximum likelihood analyses. \citet{Woodhams}  describes a technique which uses the Gaussian PSFs of individual events to identify excesses and assumes large number statistics, so that fluctuations are Gaussian via the central limit theorem (see Section \ref{WATSourceID} for further discussion of this assumption).   A similar method was used by the Fly's Eye group to analyze data from  Cygnus X-3 \citep{FlysEyeTeq}, where the method allowed them to account for the fact that their direction reconstruction was much better in one dimension than the other.  They also used Gaussian PSFs and assumed Gaussian fluctuations \citep{FlysEyeTeq2}. The assumption of Gaussian fluctuations resulted in a significance map which contained narrow significance spikes caused by the near-coincidence of pairs of well measured showers and uncharacteristic of a true source, which they accounted for using an expected point source template.

The analysis technique presented in this paper is an extension of the  Gaussian weighting techniques used by Woodhams and Cassiday et al.\ to  PSFs of arbitrary shape and to Poisson statistics. We also discuss several ways in which it can be made computationally efficient. In Section \ref{WATSkymap} the conceptual and mathematical underpinnings of the analysis are presented.  Section \ref{WATSourceID} then details source detection and the effects of various approximations, with Section \ref{WATSensitivity} comparing the sensitivity of the new analysis to optimal binned and maximum likelihood techniques.

\section{Conceptual Foundation}

\label{WATSkymap}



To build a new analysis, we must first decide what question we are trying to answer.  Are we trying to measure parameters of possible sources, such as their flux or spectrum, or are we content merely to  discover that they exist?  If we wish to estimate parameters, we may  need to parametrize the range of possibilities, for example by specifying a spectral form such as a power law.  This is the starting point for maximum likelihood searches.  If we are  willing to go further and specify our degree of belief about the number and variety of sources that may exist, we can develop a fully Bayesian analysis, as in \citep{BeyesianRef}.  Such analyses are often very computationally demanding, and thus not suitable for applications sensitive to computational constraints such as the real-time detection of transient sources. Moreover, for weak sources it may not be possible to determine parameters such as spectra with any precision, and therefore the additional complexity is not helpful.

We are interested in developing a simple and fast analysis technique for source detection. To clarify the exposition we separate the discussion of the analysis into two separate sections. In this section we concentrate on building a sky map which estimates the true photon distribution from the measured event directions and characteristics.  The next section then uses the sky map to identify sources.  For this paper we adopt a frequentist approach to the statistics because it is well suited to fast null-hypothesis tests.


We begin by considering how to represent our knowledge of the event  positions and characteristics in a sky map. The PSF can be defined as  the normalized probability density distribution for the true event position given the measured event position:
\begin{equation}  
\label{PSFdef}   
PSF(\vec{k}_t-\vec{k}_{i}) =\partial P(\vec{k}_{t}|\vec{k}_{i})/   \partial\Omega,
\end{equation}
 where $\vec{k}$ is a vector on the unit sphere, $\vec{k}_{i}$ is the  measured location, and $\vec{k}_{t}$ is the true location.\footnote{ Since in astrophysics we are only concerned with the direction of the incoming photon, it is useful to represent this direction as a vector on the unit sphere where $\Omega$ is the solid angle. }  In the definition of Equation \ref{PSFdef} we have explicitly excluded information about known source positions from the sky map.  This simplifies the construction of a sky map considerably, and known sources can be introduced in the source identification step described in Section \ref{WATSourceID}. The PSF can equivalently be defined by $\partial   P(\vec{k}_{i}|\vec{k}_{t})/\partial\Omega$, which by Bayes' formula is identical to the form in Equation \ref{PSFdef} for a uniform prior (no source information).   
 
 For many gamma-ray observatories, the width and shape of the PSF depends strongly on the individual event characteristics $\psi_i$, such as energy, incidence angle, and where the photon struck the detector.  The event characteristics can also be used to distinguish photon events from background events\footnote{ In this paper, the term ``background'' is used in two senses.  In the context of individual triggers, a ``background event'' refers to an event not caused by a gamma ray photon (i.e., from cosmic rays or instrumental effects).  In the context of searching for excesses from new sources in the sky map, however, the ``background" refers to the expected $w(\vec{k})$ distribution under the null hypothesis, and this may also include gamma rays from known sources (either diffuse or point source).  We will differentiate these two usages by reserving ``background event" for individual events and ``background" for the sky map expected under the null hypothesis.} and determine $P_{\gamma}(\psi_i)$ --- the probability that the event was a photon rather than a background  event.  The PSF can then be multiplied by $P_\gamma$ to create a distribution of the photon probability density for the $i^{\rm th}$ event being a real photon coming from some direction $\vec{k}$:
\begin{equation}  
\label{probDensityEq}   
p_i(\vec{k}) = p(\vec{k}|\vec{k}_i,\psi_i) = PSF(\vec{k}-\vec{k}_{i},   \psi_i)P_{\gamma}(\psi_i).
\end{equation}

\begin{figure}  
\begin{center}    
\includegraphics[height=6.75in]{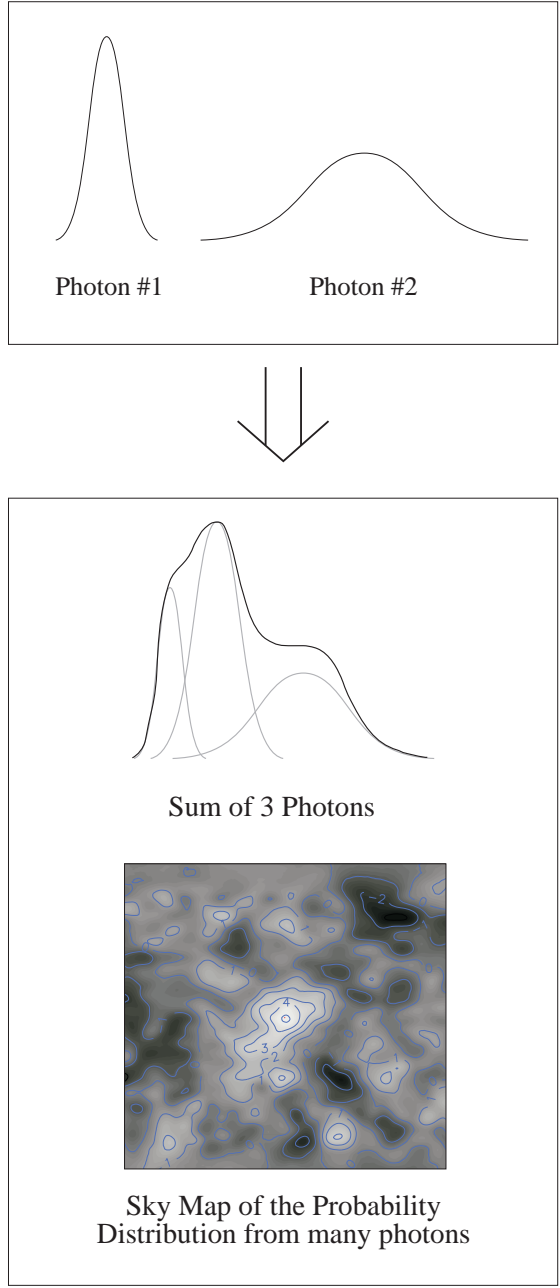}    
\caption{This diagram shows how individual photons are added to form a sky map in this analysis.  In the top frame are two events with PSFs determined by the event characteristics.  The lower frame shows how events are then added together to form a photon density map, first as a one dimensional example of three events, then a more realistic two dimensional example incorporating many events.}    
\label{PSFSummationDiagram}  
\end{center}

\end{figure}

A sky map can be created by adding together the photon probability distributions of many events to form an overall map and form an estimate of the total photon density $w$, where 
\begin{equation}  
\label{weightSum:Eq}   
w(\vec{k})=\sum_i^ {\text{all showers}} p_i(\vec{k})=\sum_i^{\text{all        showers}}PSF(\vec{k}-\vec{k}_{i},\psi_i)P_{\gamma}(\psi_i).
\end{equation}
  Figure \ref{PSFSummationDiagram} provides a graphical description of this procedure.  The resulting sky map is equivalent to the estimated total photon density assuming a uniform prior event distribution.

The continuous sky map implied by Equation \ref{weightSum:Eq} does not map efficiently to a computer algorithm due to the infinite number of points $\vec{k}$.   However, the spatial scale for fluctuations across the sky map is set by the width of the narrowest PSF. The sky map can be digitized by sampling the estimated total photon density $w(\vec{k})$ at individual  points on the map surface.  If the spacing of these samples is small  compared to the narrowest PSF, the information loss can be made arbitrarily small.  In practice, we construct digitized templates for the various cases of the PSF and add them point by point to form a digitized sky map.  The resulting algorithm can very rapidly build sky maps of the estimated photon probability density.

There are several additional features of this digitized sky map which are useful. Because the value at a location on the sky map represents the estimated photon density at that point and is not an integral over nearby locations (as is common in binned analyses), the spacing between the points need not be uniform and tiling problems associated with binning a spherical sky are avoided. Additionally, two sky maps which share a sampling pattern can be summed. An 80-second sky map can be formed by adding two sky maps of 40 seconds' duration --- a significant computational advantage when hunting over multiple time scales.

Spectral information has been ignored in this discussion, but can be included in a completely analogous manner. The key is to determine the normalized energy probability density function $\partial  P(E_t|E_i,\psi_i)/\partial E$ --- the one dimensional energy analog of the PSF. This energy distribution can then be multiplied by $P_\gamma(\psi_i)$ to form the estimated photon energy probability density $p_i(E)$ and added as a third independent axis of the sky map.   The resulting three dimensional sky map is harder to visualize, but again represents the total estimated photon density distribution. Similarly, the probability density of any other parameter of interest (such as polarization) can be added to a multidimensional sky map to enrich the representation of the data.  In conclusion, we can form a digitized sky map that represents our direct knowledge by summing the photon probability density distributions of each event and digitizing the resulting map.

\section{Source Detection}
\label{WATSourceID}

Now that we have a sky map, we need to tackle the issue of source detection in the low signal-to-noise and photon counting regime of most gamma-ray observatories. In a discovery mode search the relevant  statistic is the probability of the background producing the observed candidate signal.  For point sources we can look at each of the sampled locations independently and ask, ``What is the  probability of the background producing the observed photon density  $w(\vec{k})$ at that location?" To answer this question, we need to  know the expected distribution $g(w | N)$ of the total photon density $w(\vec{k})$, which depends on the number of events $N$ added to the sky map.\footnote{\label{footnote3}Formally $g$ is the expected probability distribution of $w$.  However, using standard notation and phrasing leads to awkward sentences like ``the expected probability distribution of the estimated photon probability density.''  In an effort to clarify the discussion, we use $g$ to refer to the expected statistical distribution of the estimated photon probability density $w$, and refer to this as the ``distribution of $w$.''  Also note that we are interested in parent statistical distributions, not the observed distributions. A potential source of confusion is the equivalence of the expected probability distribution of $w$ at any given point and the distribution of $w$ over points $\vec{k}$ on the sky; i.e., the probability that a random point $\vec{k}$ will have a given value of $w$ is also the expectation value for the fraction of points $\vec{k}$ with similar event characteristics having that value.  We generally will not distinguish between these two equivalent applications of the parent distribution.
%
}
If the distribution $g$ is Gaussian, then knowing the average and RMS is sufficient to determine the probability of the background producing an  observed photon density.  However, for typical PSFs the expected distribution of $w$ for a single event is far from Gaussian, and one must sample from the distribution  --- i.e., add events to the sky map ---  a surprisingly large number of times before the fluctuations on the sky map are well described by a Gaussian.

\begin{figure}  
\begin{center}    
\includegraphics[width=5.75in]{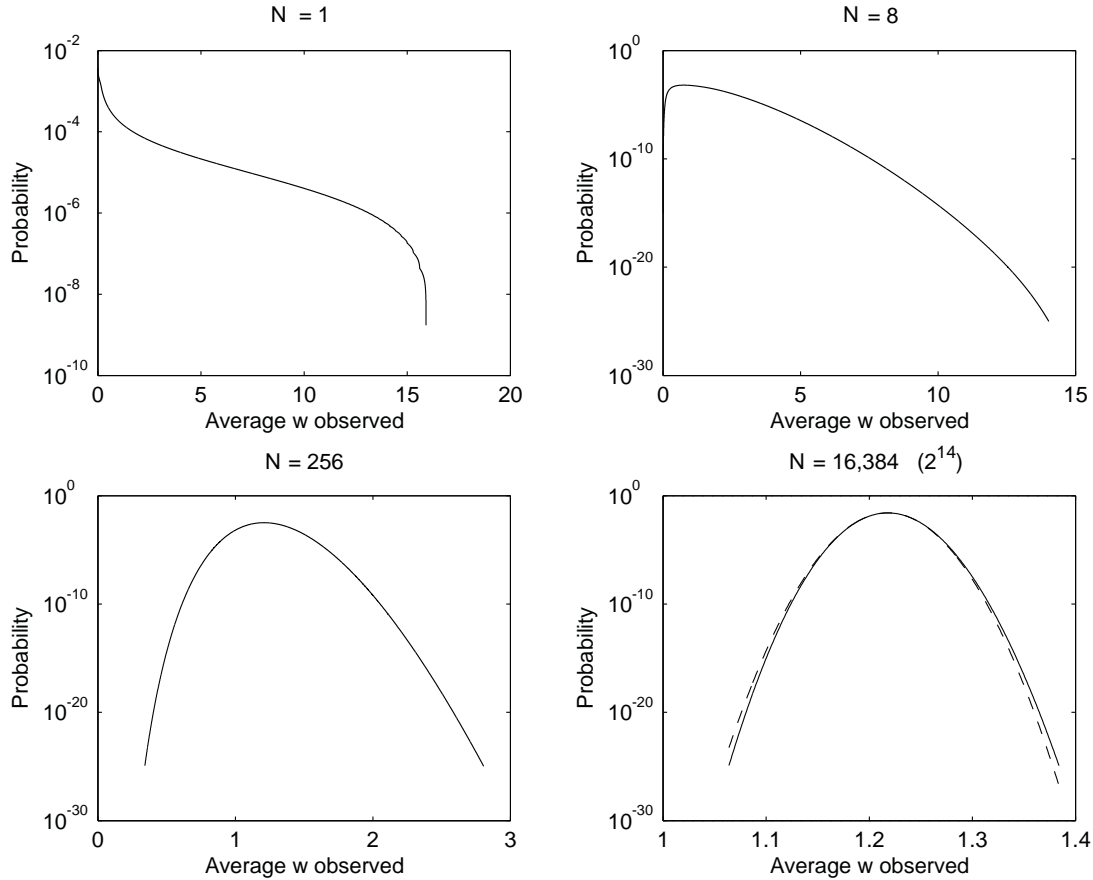}    \caption{These plots illustrate a typical photon density distribution created by events with a range of PSF widths.  In this example the events have 2 dimensional Gaussian PSFs, with the width of the PSFs equally distributed between $\sigma/10$ and $1\sigma$. The top left panel shows the distribution of $\overline{w}$ expected from a single event randomly placed on the sky.  This distribution is highly skewed due to the large number of small values in the tails of the Gaussian PSFs (the maximum $\overline{w}\approx15.9$ is given by the peak value of a normalized $\sigma/10$ Gaussian).  The other panels show the photon density distribution $g(\overline{w}|N)$ for three values of $N$ and a smooth background as the distribution becomes increasingly Gaussian in shape. Note that the range shrinks significantly from one plot to the next. In the last panel the dashed line indicates the best-fit Gaussian distribution.  $N$ must be very large before the distribution becomes Gaussian due to the highly skewed nature of the $N=1$ distribution.  Even for the largest $N$ case shown, the Gaussian approximation substantially overestimates the significance of fluctuations to large $\overline{w}$.}    
\label{weightProb}  
\end{center}

\end{figure}

To emphasize the differences in the shape of the $w$ distribution, we change variables from the total photon density $w$ to  $\overline{w} \equiv w/N$, the normalized photon density.\footnote{ Since the PSFs cover the sky, $N$ does not vary from point to point on the sky map. Section \ref{TruncatedPSF} discusses source identification with the approximation of local PSFs.} We now consider the expected distribution of $\overline{w}$ for a single event, which can then be used to calculate the expected distribution of  $\overline{w}$ for a sky map containing any number of events. Since the photon density added to a location on the sky map varies with the  angular separation $\vec{k}-\vec{k_i}$, there is a distribution of photon densities $g(\overline{w}| N=1)$ which a single background event may add to any given sky map location.  (Recall that the probability  distribution at any single point is equivalent to the expected distribution of $\overline{w}(\vec{k})$ over similar points on the sky as explained in Footnote \ref{footnote3}.) The  probability of the background producing a given observed photon density is defined by this distribution of photon densities, and since the distribution may not be Gaussian we must determine the complete  distribution of $\overline{w}$ and not just the expected average and RMS of $\overline{w}$.  Two-dimensional Gaussian-like PSFs have a large area for which $w$ is small, so the photon density distribution for a single event is highly skewed with low photon densities being much more common than high photon densities (see Figure \ref{weightProb}, panel 1).  The distribution observed in a specific detector will depend on the PSF characteristics and can be determined theoretically or measured directly. 

As the number of events added to the sky map increases, the expected  distribution of the total photon density $g(\overline{w}|N>1)$ evolves, eventually approaching a Gaussian distribution.   The $N > 1$  photon density spectra can be directly calculated by repeatedly convolving the $g(\overline{w}| N=1)$ distribution with itself.  Some  example distributions for a smooth background are shown in Figure \ref{weightProb}.  Note that  while $g(\overline{w}| N)$ becomes Gaussian for large $N$, the convergence is very slow due to the highly skewed $g(\overline{w}| N=1)$ distribution.  For traditional Gaussian weighting techniques the number of events needed before large number statistics are appropriate can be larger than one would na\"{i}vely predict.

The $g(\overline{w}|N)$ photon density spectra define the statistical  variations of $\overline{w}$ observed in the sky map.  Since the distribution of event characteristics can vary with location, the photon density spectra are more correctly written as  $g(\overline{w}|N,\vec{k})$ to include this spatial variation.  The  probability of fluctuations of the background producing an  observation with a photon density greater than or equal to the observed value is then given by the integral of the normalized photon density distribution $g(\overline{w}|N,\vec{k})$. The resulting integral probability
\begin{equation}  
\label{baseProb}   P(\overline{w}\ge\overline{w}_{obs}|g(\overline{w}|N,\vec{k})) = \int_{\overline{w}_{obs}}^{\infty} g(\overline{w}|N,\vec{k})\, d\overline{w}
\end{equation}
 defines the probability of the background fluctuations producing such an observation.

In general the total number of events $N$ added to the sky map may experience small Poisson fluctuations around the expected background  $N_{exp}$, adding an additional variation to the sky map which is not  included in Equation \ref{baseProb}. It is common in gamma-ray astronomy to use the total number of events observed across the field-of-view to determine the expected background $N_{exp}$, so that  $N\equiv N_{exp}$ and the probability formulation in Equation \ref{baseProb} is exact. Fluctuations in $N$ can be included by using the formalism developed in Section \ref{TruncatedPSF}. However, if the PSFs are narrow compared to the field-of-view, the variations on the sky map are typically dominated by fluctuations in the average photon density $\overline{w}$ and Equation \ref{baseProb} is appropriate.  In cases where wide PSFs or very low expected background rates make the global event rate a significant variable, the formalism developed in the next  subsection should be used instead.

\subsection{Signal Detection with Truncated PSFs}
\label{TruncatedPSF}

The PSFs used to generate the sky map need not match the true PSFs. The PSF analogs used in the analysis are called ``weighting functions" to differentiate them from the true PSFs, and in many applications it is convenient to truncate the weighting functions at some angular diameter.  Truncating the weighting functions can significantly improve the computational speed of the analysis since only sky map locations near the position of a new event must be updated, not the entire sky map.  The cost of truncating the weighting functions is a further complication of the statistics.  When the weighting functions are truncated the number of events $N$ summed at a location on the sky map may experience strong Poisson fluctuations around the expected background value $N_{exp}$, making $N$ an important part of the probability calculation (where we now take $N = N(\vec{k})$ to be the local number of events summed to form $\overline{w}(\vec{k}) = w(\vec{k})/N$, rather than the total number of events added to the sky map). The expected distribution $g$ now becomes a two dimensional probability distribution $g(\overline{w},N)$.  We will reserve $g(\overline{w}|N)$ for the one dimensional distribution of the previous section and use $g(\overline{w},N)$ to indicate the two dimensional distribution which arises when $N$ varies due to PSF truncation.\footnote{\label{1d2dfootnote}The one dimensional $g(\overline{w}|N)$ can also be viewed as a slice of the two dimensional $g(\overline{w},N)$ at a given $N$ value.}  The question of how to determine the correct probability when the probability density is a function of two independent variables can be answered using an argument analogous to the one used by \citet{FeldmanCousins} in their paper on confidence limits.

\begin{figure}  
\begin{center}    
\includegraphics[width=5.75in]{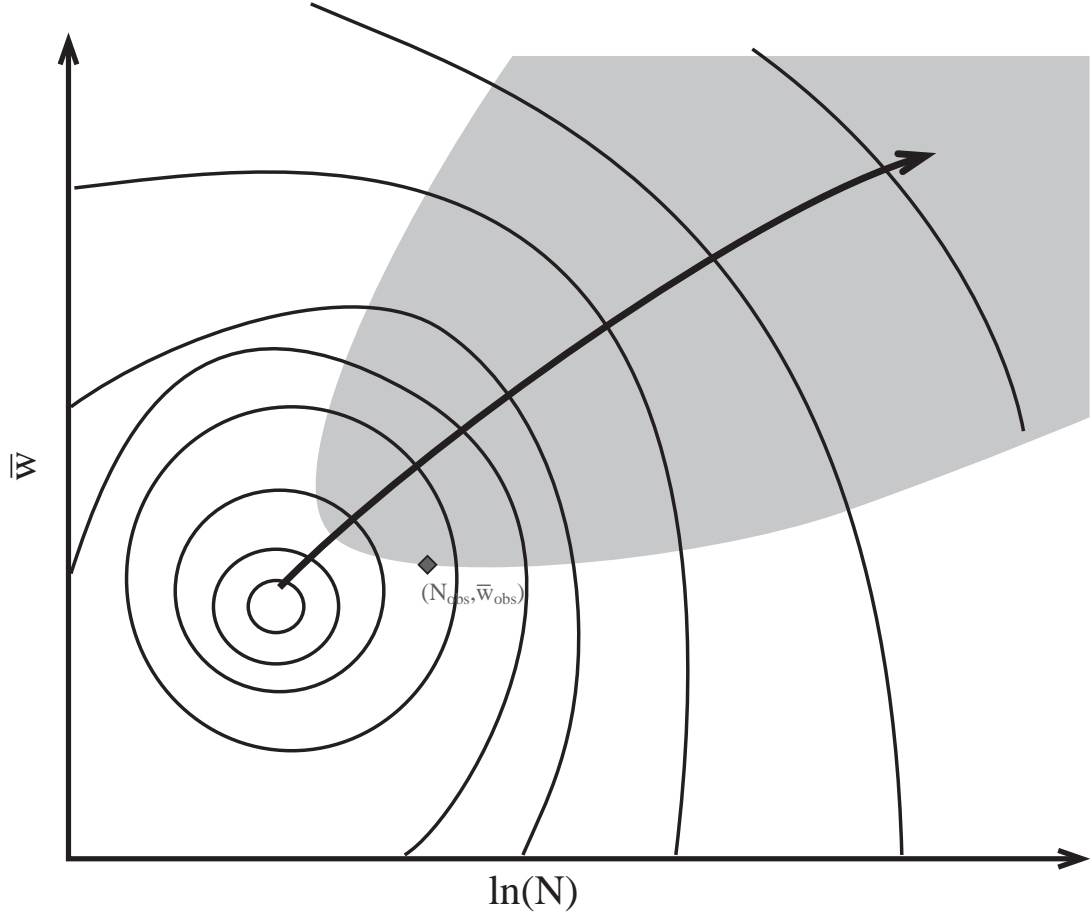}    \caption[Determining the True Probability]{A cartoon of how to determine the probability of the background producing an event which is more signal-like than an observation.  The contour lines are the probability density of the background $g(\overline{w},N)$ producing a particular combination of $\overline{w}$ and $N$;  the heavy line indicates the trend where signal events are likely to be.  The diamond is a particular observation, with the shaded region including all of the points where the likelihood ratio is equal to or more signal-like than the current observation.}    \label{ProbabilityDiagram1}  
\end{center}
\end{figure}

The probability density distribution of the background producing the observation at a given point $\vec{k}$ as a function of $\overline{w}$ and $N$ is represented by the contour lines in Figure \ref{ProbabilityDiagram1}.  The distribution in Figure \ref{ProbabilityDiagram1} is the two dimensional expected distribution of $g(\overline{w},N)$ as a function of $\overline{w}$ and $N$, and is analogous to the one dimensional $g(\overline{w}|N)$ distribution from the previous section. Given a theoretical model of a possible point source, a similar probability density distribution can be constructed for the expected signal. The ratio of the likelihood of observing the given $\overline{w}$ and $N$ assuming the null hypothesis (i.e., only background) to the likelihood assuming both the background and the signal contribute can then be determined for every point in the plane.  This point-by-point likelihood ratio is analogous to Equation 4.1 in \citet{FeldmanCousins}. Feldman and Cousins used the likelihood ratio to define how to order points from most signal-like to least, then determined a confidence interval by adding points until a predetermined probability was reached ({\it e.\!\,g.},\ 90\%). Here we can use the likelihood-ratio ordering to determine which values of $\overline{w}$ and $N$ are equally or more signal-like than the current observation.  The probability of the background producing an event which is more signal-like is then determined by integrating the probability density of the background $g(\overline{w},N)$ at all points where the likelihood ratio of background to signal  plus background is  equal to or lower (more signal-like) than the likelihood ratio at the observed location.

This process is shown graphically in Figure \ref{ProbabilityDiagram1}.  
 In this example signal events tend to have both higher $\overline{w}$ and $N$ as indicated by the heavy line.  An observation is indicated by the diamond and the shaded region includes all of the positions where the likelihood ratio is equally or more signal-like than the observed position.  The probability of the background producing an event which is more signal-like is then given by integrating the background probability density $g(\overline{w},N)$ in the shaded region.  This defines the  probability of the background producing an observation when there are two (or more) observed variables.

An equivalent description of the probability determination views the  likelihood ratio as a prescription for mapping the two dimensional  $g(\overline{w},N)$ probability distribution to a one dimensional  probability distribution, where the likelihood ratio serves as the  independent variable.  The probability of the background producing an  event equally or more signal-like is then the integral of the one dimensional probability distribution for all likelihood ratios more  signal-like than the observation, analogous to Equation \ref{baseProb} in the previous section.  

When implementing this analysis the two dimensional integrals over  $\overline{w}$ and $N$ can be performed in advance, enabling very fast  source identification.  Calculating the probabilities involves modeling the expected signal; determining the likelihood-based integration regions; and forming a table in $\overline{w}_{obs}$ and  $N_{obs}$ of the integral background probability densities.  This table can then be used to quickly look up the correct probability for a given signal, requiring no on-the-fly probability calculations.

\subsection{Further Source Detection Approximations}

\label{FurtherApprox}

\begin{figure}  
\begin{center}    
\includegraphics[width=5.75in] {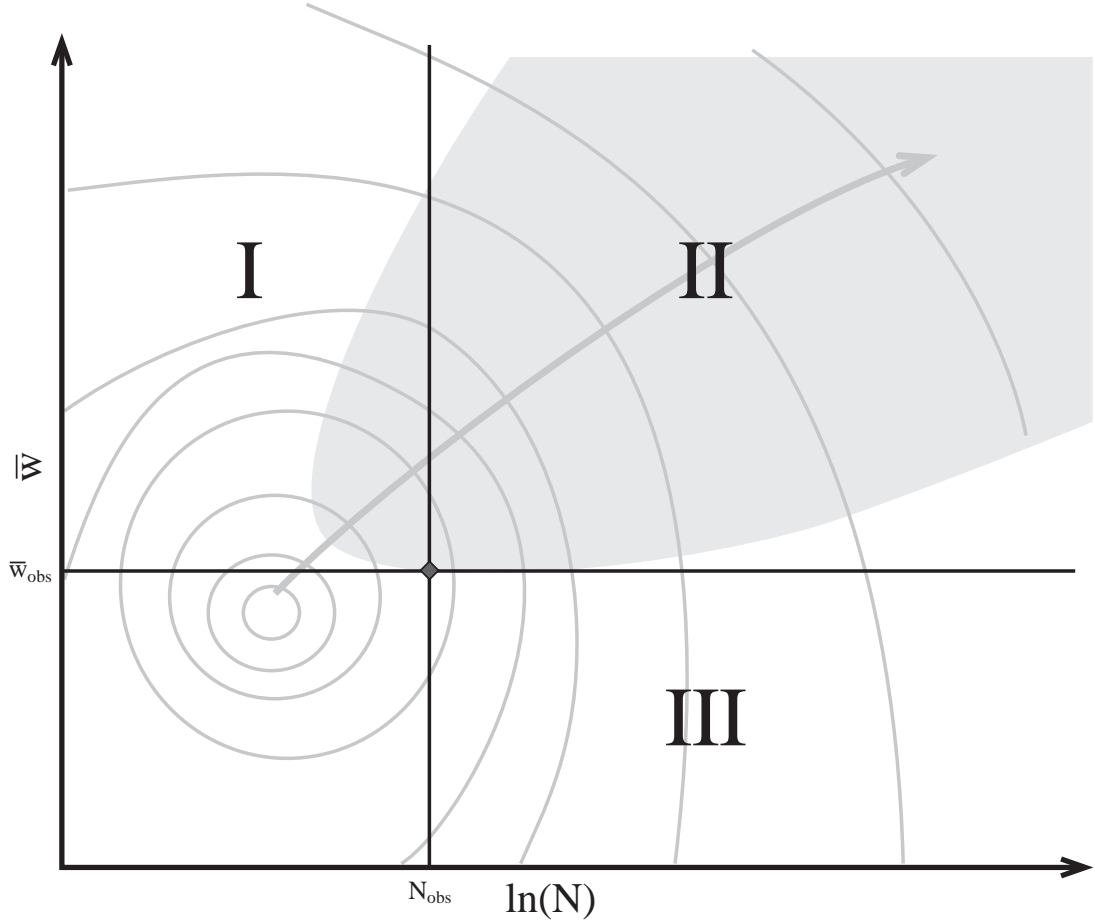}
\caption[Approximation to the True Probability]{This is the same diagram as Figure \ref{ProbabilityDiagram1}, with lines indicating the $\overline{w}_{obs}$ and $N_{obs}$ values.  Equation \ref{CorrectProbEq} approximates the true probability by integrating the background probability density in region II.}    
\label{ProbabilityDiagram2}  
\end{center}
\end{figure}

In the case of Milagro, source identification is complicated by the lack of a reliable source model, making any determination of the  likelihood ratio uncertain.  Without a model, the question becomes how can an appropriate integration region be chosen? Since we expect a signal to increase both the measured $\overline{w}_{obs}$ and $N_{obs}$ values, the simplest approximation to the true probability of the background producing an event as signal-like or more than the current observation is the integral of the background probability density for all $\overline{w}$ and $N$ greater than the observed value, as  depicted by region II in Figure \ref{ProbabilityDiagram2}.  
The resulting approximate probability of the background producing the observed signal is given by
\begin{equation}  
\label{CorrectProbEq}  
\sum_{N=N_{obs}}^\infty P(\overline{w}\ge   \overline{w}_{obs}|g(\overline{w}|N,\vec{k}))P(N|N_{exp}),
\end{equation} 
where we have used the relationship from Footnote \ref{1d2dfootnote} to change from the two dimensional $g(\overline{w},N)$ to the one dimensional $g(\overline{w}|N)$.  If the background probability density distribution $g(\overline{w}, N,\vec{k})$ changes slowly with $N$ ($\partial P(\overline{w}\ge   \overline{w}_{obs} | g(\overline{w}|N,\vec{k})) /\partial N$ $\ll$ $  \partial P(N|N_{exp})/\partial N$), Equation \ref{CorrectProbEq} can   be further approximated by separating the probability terms  
\begin{equation}    
\label{Prob:Eq}     
P(\overline{w}\ge    \overline{w}_{obs}|g(\overline{w}|N_{obs},\vec{k}))     P(N\ge N_{obs}|N_{exp}),   
\end{equation}
significantly reducing the memory footprint of the lookup tables.   This approximation was used for the 40 s -- 3 hour transient search performed by \citet{MyThesis} for the Milagro observatory. 

Any approximation to the correct (but unknown) region of integration would be expected to degrade the sensitivity of the analysis. However, there is an additional complication that arises with this particular choice of integration rule because region II does not contain all points having a total probability from Equation 5 equal to or smaller than the observed value.  Hence the sum systematically underestimates the probability (overestimates the significance) of obtaining the particular observation as a fluctuation of the background, for the following reason.

When using the probability integrals defined by the likelihood ratio (see Section \ref{TruncatedPSF}) there is a one-to-one mapping between the integration region and the value of the integral, i.e., the total  probability of the background producing a fluctuation equally or more  signal-like than the observation.  This can be seen in Figure  \ref{ProbabilityDiagram1} where all of the events with the same likelihood ratio as the current observation lie along the edge of the  integration region, and their corresponding probabilities would be determined by integrating exactly the same region.  We can continue this argument and break the entire plane into a set of infinitesimally thin contours where all of the events on each contour have the same likelihood ratio and share the same integration region and total probability. This also means that the integration region for an observation contains all events with an equal or lower probability of being produced by the background, since the contours for the lower probability events are contained by the contours for the higher probability events.  This can be visualized for smooth likelihood distributions as a set of concentric contours, with each contour just surrounding the contour with a slightly lower likelihood ratio and total probability.\footnote{ This containment effect can also be obtained from the mapping argument used in Section \ref{TruncatedPSF}, and holds when the integration regions are not contiguous or the likelihood ratio   distribution is discontinuous in $\overline{w}$ and $N$.}

\begin{figure}  
\begin{center}    
\includegraphics[width=5.75in] {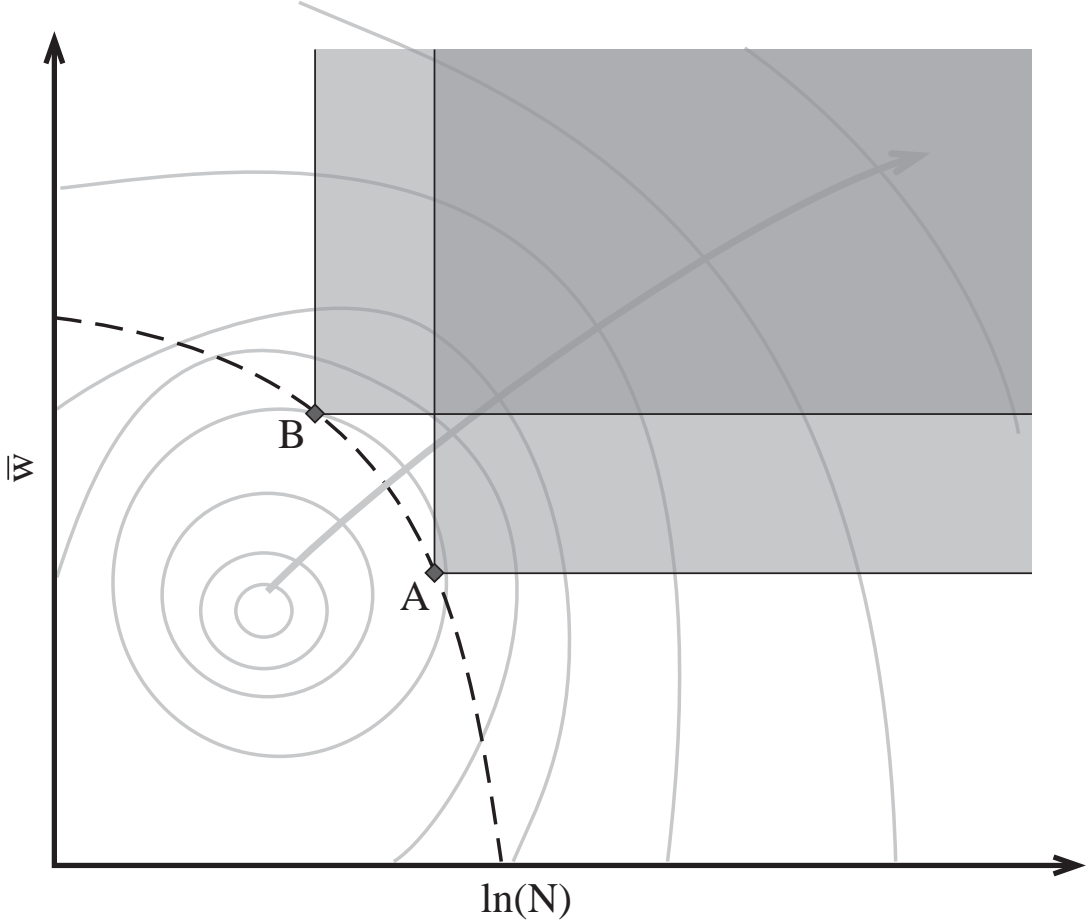}    \caption{In this example, the probability integrals for       observations A and B both have the same value using the       approximation from Equation \ref{CorrectProbEq}.  The set of all such points, illustrated by the dashed line, forms a contour of equal total probability. The probability of seeing an observation equally or more signal-like than observation A (an observation with equal or lower probability) is systematically under-estimated because observation B and portions of the associated integral are not included in the integral of A.  This leads to distortions in the probability histograms produced when using this approximation.}    
\label{ProbabilityDiagram3}  
\end{center}

\end{figure}

However, with the rectangular integration regions of the approximation in Equation \ref{CorrectProbEq}, the contours of equal integrated probability are no longer identical to the regions of integration.  As shown in Figure \ref{ProbabilityDiagram3}, multiple observations may produce equal integrals without integrating the same region.  These observations may be thought of as forming a contour of equal calculated probability, which is distinct from the regions integrated over for any given point.\footnote{ Because the integration regions are rectangular, the contour (thought of as a relation between $\overline{w}$ and $\ln{N}$) must be a monotonically decreasing function as sketched in Figure 5.}   Note that the region which is integrated to estimate the probability is now smaller than the region enclosed by the contour of equal estimated probability.  This means that the probability of obtaining a background fluctuation as signal-like as the current observation (given by the integral of the region to the right of the contour) is actually larger than the estimated value (given by the integral over the rectangular region).  This approximation thus slightly distorts the observed  probability histograms.  The distortion can be  corrected for using either the background measurements of the detector or  Monte Carlo simulations \citep{MyThesis}.  While not a critical issue, this distortion is an annoyance and it is hoped that a better integration rule can be developed which would more closely approximate the correct region of integration and thus reduce the distortion of the probability histograms.

It is interesting to note that a binned analysis can be obtained by  integrating regions II and III of  Figure \ref{ProbabilityDiagram2}. An integral of all $N$ greater than the observation is equivalent to a  binned analysis with a bin-size equal to the truncation distance applied to the PSF, and  recreates the familiar integral Poisson distribution.

In conclusion, sensitive source identification can be performed in several ways, depending on the requirements of the analysis.  If the true all-sky PSFs are used and the total number of events detected is not an informative parameter, then Equation \ref{baseProb} is appropriate.  If the weighting functions are truncated to increase the speed of building sky maps, or if the total number of events in the sky map is important, then the probability can be determined using  likelihood ratio ordering and tabulated in advance to give a very fast analysis with little sensitivity degradation (Section \ref{TruncatedPSF}). This can be further approximated at the cost of some signal sensitivity as described in this subsection.

\section{Sensitivity}
\label{WATSensitivity}

We would like to compare the sensitivity of this analysis technique to  maximum likelihood and binned analyses. Unfortunately, there is no simple analytic way of comparing the various implementations of maximum likelihood to either the analysis presented in this paper or the optimal binned analysis, and a Monte Carlo simulation tailored to the particular application must be used. 

In general, comparisons between the analysis method presented here and an optimal binned analysis also require application-specific Monte Carlo simulations.  However, we can illustrate the key differences using a few toy models with analytic solutions. In the limit of large  statistics (see Section \ref{WATSourceID}), we can obtain analytic solutions for this analysis and an optimal binned analysis for a detector with a single Gaussian PSF, and for a detector with events drawn from two Gaussian PSFs of different widths. In this limit of large statistics and Gaussian PSFs, and also excluding weighting for photon probability $P_\gamma$, the analysis presented in this paper becomes identical to the Gaussian weighting analyses of \citet{Woodhams} and  \citet{FlysEyeTeq2}. Thus all the results in this section may also be  applied to traditional Gaussian weighting techniques. 

For these toy models, $N_s$ is the number of signal photons and $b$ is the number of background events per unit solid angle near the signal. After background subtraction, the significance of the signal is given by the signal/noise ratio and has the form $AN_s/\sqrt{b}$, where $A$ characterizes the sensitivity of the search and is the object of the following calculations. Because the PSFs in these models are azimuthally symmetric, $\vec{k}-\vec{k}_i$ depends only on the angular separation between the source location and the reconstructed event location.  To simplify the equations, $r$ is used to denote the angular separation between the source ($\vec{k}$) and reconstructed positions  ($\vec{k}_i$).  Furthermore, we ignore the curvature and finite size of the celestial sphere, letting r range from 0 to infinity, and rely on the exponential decay of the PSF to effectively cut off the integrals.

For a single Gaussian, the signal observed in a circular bin of radius $R$ is given by the integral of the PSF 
\begin{equation}  
\label{ }   Signal=\int_0^R\frac{N_s}{2\pi\sigma^2}e^{-r^2/2\sigma^2}2\pi r\, dr    = N_s(1-e^{-R^2/2\sigma^2}),
\end{equation}
 and the noise is given by the square root of the number of background  events $\sqrt{b\pi R^2}$.  This ratio is maximized for $R=1.585\sigma$, giving a sensitivity parameter A of $0.255/\sigma$ for an optimal binned analysis in a detector with a single Gaussian PSF. 

For the analysis presented in this paper, the signal from a point source with $N_s$ photons is the probability distribution of the photon positions (the true point spread function $PSF$) times the weight given to each photon (the weighting function $PSF'$):
\begin{equation}  
\label{WATSignal}   Signal =\int_0^{\infty} N_s\ PSF\ PSF'\ 2\pi r\, dr.
\end{equation}
 Since the weighting function and the PSF are the same Gaussian function in this example, the integral becomes
\begin{equation}  
\label{ }   Signal =\int_0^{\infty} N_s\Bigl[\frac{1}{2\pi\sigma^2}     e^{-r^2/2\sigma^2}\Bigr]^2 2\pi r\, dr =\frac{N_s}{4\pi\sigma^2}.
\end{equation}
 The noise is given by the square root of the variance of the probability density.  In the limit of large statistics, the variance is given by integrating the flat background distribution times the square of the weighting function:
\begin{equation}  
\label{WATNoise}   Noise =\Bigl[\int_0^{\infty} b\ PSF'^2\ 2\pi r\, dr\Bigr]^{1/2}.
\end{equation}
 Since the weighting function is the same Gaussian as the true PSF, this is the same integral as used  for the signal with $b$ replacing $N_s$. The signal to noise ratio becomes $\frac{N_s}{\sqrt{4\pi\sigma^2}\sqrt{b}}$, giving a sensitivity parameter A of $0.282/\sigma$.  This implies that the new analysis is  $\sim10\%$ more sensitive than an optimal binned analysis for a detector with a single Gaussian PSF.  This matches the results obtained by \citet{Woodhams} for the Gaussian analysis in the same limits.  Woodhams argued that this 10\% improvement should be a lower limit, and that detectors which have a distribution of PSFs should benefit even more from a weighted analysis. 

The next toy model has two Gaussian PSFs, with 25\% of the events coming from a PSF of width 0.33$\sigma$, and 75\% from a PSF of width  1$\sigma$. From the characteristics of each event, we are able to assign it to one of the two PSF groups.  Following the previous calculation, the optimal bin size is $0.764\sigma$ and the sensitivity  parameter is $0.312/\sigma$ for the optimal binned analysis.  For the  new analysis the sensitivity parameter is $0.489/\sigma$, or a $\sim 56\%$ improvement in sensitivity over the binned analysis.  Hence, there are situations where the more sophisticated approach yields  substantial improvements in signal sensitivity. 

However, the improvement depends very much on the distribution of PSFs, and in special circumstances the improvement can be zero. To show that the 10\% improvement from a single Gaussian PSF is not a lower limit as argued by \citet{Woodhams}, consider a distribution of  PSFs given by $s(\sigma)$.  The general problem of finding the signal in a round bin becomes
\begin{equation}  
\label{ }   N_s\iint_0^R PSF(\sigma,r)s(\sigma)2\pi r\, dr\, d\sigma,
\end{equation}
 and the signal in the analysis presented here becomes 
\begin{equation}  
\label{WATSignalSpectrum}   N_s\iint PSF(\sigma,r) PSF'(\sigma,r) s(\sigma)2\pi r\, dr\, d\sigma.
\end{equation}
 For a flat distribution of Gaussian PSFs from width 0.1$\sigma$ to width 1$\sigma$, the new analysis gives less than a 7\% improvement over the binned analysis despite the wide range of PSFs used. In retrospect, this can be explained by reversing the order of integration. By integrating the distribution of PSFs first (over $d\sigma$), a composite PSF can be obtained which has a distinctly  non-Gaussian profile. By choosing the appropriate PSF and distribution, a composite PSF with a top-hat profile could be generated, and in this extreme case the optimal binned analysis would be just as effective as the analysis presented in this paper. This can be seen by realizing that the new analysis technique with a top-hat  weighting function
\begin{equation}  
\label{tophat:Eq}   PSF'(\vec{r}) =\begin{cases}     a &\vec{r}<R\\     0 &\vec{r}\ge R\ ,  \end{cases}
\end{equation}
 where $R$ is the size of the bin and $a$ is a constant, is identical to a binned analysis.  Referring to the probability of a background  fluctuation producing the observed signal as described in Figure \ref{ProbabilityDiagram1} and Section \ref{TruncatedPSF}, the top-hat  weighting function leads to $\overline{w}= a$,  since all of the events with a non-zero probability density have a  probability density of $a$.  The total probability is then solely determined by the probability of the background producing an  observation with $N$ greater than $N_{obs}$.  This is simply the Poisson probability of seeing $N>N_{obs}$ events inside a bin of radius $R$ --- exactly the same result as a binned analysis (this can  also be seen by substituting the top-hat weighting function into  Equations \ref{WATSignal}, \ref{WATNoise} and \ref{WATSignalSpectrum}).  It can also be shown that the optimal weighting function to use in either the analysis presented here or Gaussian weighting is the true PSF \citep{Woodhams}. Since the optimal  weighting function is the true PSF, and the new analysis with a top-hat weighting function is identical to a binned analysis, it follows that the sensitivity of the analysis presented here is never worse than a binned analysis, and would only be equal for a detector with a top-hat composite PSF.  In general, the less square the composite PSF is, the more effective the new analysis will be.

One final topic we can explore with simple examples is the sensitivity of these analyses to errors in the expected signal characteristics.   Returning to the example with two Gaussian PSFs, we can compare the  sensitivity of both analyses to signals where all the signal events come from either the narrow or wide PSFs while the expectation is still for a 25\% -- 75\% division between the PSFs. In these examples,  the bin size or background distributions will be wrong, and we can explore how errors in the expected PSF affect the sensitivity of the  analyses.  If the PSF of the signal is $0.33\sigma$ (all narrow PSF events), the analysis presented here is more than twice as sensitive as the binned analysis (114\% improvement).  At the opposite extreme, if the PSF of the signal is $1\sigma$ (all wide PSF events), then the binned analysis is nearly 13\% more effective than the new analysis  technique.  This surprising result comes about because the power of the new analysis technique comes from weighting the events with the expected PSF.  However, if the expected PSF is wrong, there can be times when the expected optimal bin/top-hat weighting function of a binned analysis happens to be more accurate than the expected PSF of the new  analysis.  This shows that there is some model dependence in the analysis presented here which can be detrimental in certain specific  scenarios.

In the preceding examples the $P_\gamma$ term from Equation \ref{probDensityEq} has been set equal to one.  This is equivalent  to a hard background cut which treats all events passing the cut  identically.  The analysis presented here can use a variable $P_\gamma$ value instead of a hard cut, and this will magnify the sensitivity advantage over a binned analysis. This can be seen by observing that a  background cut is equivalent to a one dimensional bin in the cut parameter, and the same argument which showed that the sensitivity of the new analysis method is greater than or equal to that of the binned  analysis applies.  In effect, background rejection adds a third dimension to the space in which photons from a signal are localized, with two dimensions corresponding to the event direction and one  corresponding to the probability of the event being a gamma ray. An  optimal binned analysis with a background cut uses a step-like probability distribution in all three dimensions, whereas this analysis uses the expected probability distributions.

In general, the sensitivity of two analyses can only be compared using an application-specific Monte Carlo simulation. There are a number of  subtleties which have been masked by the simplicity of these examples,  including the effect of fluctuations (on all parameters) in the limit of low statistics. For GRB searches, the limit of large statistics does not hold and the similarity between Gaussian weighting and the  analysis presented in this paper is broken. Gaussian weighting as developed by \citet{Woodhams} can only be used in the limit of very large statistics, and the analysis technique developed here can be seen as an extension of Gaussian weighting to the regime of Poisson  statistics and PSFs of arbitrary shape. \citet{CygnusTeq} performed a  Monte Carlo simulation for the simple case of a single Gaussian PSF in the limit of large statistics, and they observed a $\sim$10\% improvement in the sensitivity of a maximum likelihood analysis over an optimal binned analysis.  This compares with the $\sim$10\% improvement obtained by the analysis presented here over an optimal  binned analysis in the same limit, as described in the first example  of this section.  This similarity implies that the analysis presented  here is similar in sensitivity to well-implemented maximum likelihood  techniques in this limit. The new analysis method is more sensitive than the binned analysis for much but not all of the possible phase  space, and should approach the sensitivity of well-implemented maximum  likelihood searches for at least some of the phase space.

\section{Conclusion}

This paper generalizes the ideas of the Gaussian weighting analysis to  point-spread functions of arbitrary shape and the regime of Poisson  statistics. A typical application of this analysis consists of three primary steps:
\begin{enumerate}
  \item \textbf{Forming the Sky Map --- Section \ref{WATSkymap}.} \textit{Repeated for each event.} Reconstruct the arrival direction of the event and determine the appropriate point spread function and $P_{\gamma}$ value using the event characteristics.  The resulting photon probability density $p_i(\vec{k})$ is then added to the sky map to form the estimated photon probability density $w(\vec{k})$ (Equations \ref{probDensityEq} and \ref{weightSum:Eq}).  The digitized PSF distributions can be tabulated to make adding events to the sky map very fast.
  
    \item \label{DistStep} \textbf{Calculating Expected Distributions of $\overline{w}$ and $N$ --- Section \ref{WATSourceID}.} \textit{Only needs to be recalculated for significant changes in the detector performance or analysis system, typically calculated offline once.}  The simplest method starts with the expected $g(w|N = 1,\vec{k})$ distribution for a single event.  This distribution depends on the weighting function $PSF'$ used by the analysis, the expected distribution of $P_{\gamma}$, and the spatial distribution of the background, and can be directly measured or calculated from Monte Carlo simulations.  The $g(w|N = 1,\vec{k})$ distribution is then convolved with itself to form the expected distribution for any number $N$ as demonstrated in Section \ref{TruncatedPSF}.  For a simple analysis as described by Equation \ref{baseProb} this is sufficient.  For the more complex analysis versions detailed in Sections \ref{TruncatedPSF} and \ref{FurtherApprox}, this is combined with the Poisson probability of $N$ to create the two dimensional expected background distribution $g(\overline{w},N)$.  Using the methods outlined in Sections \ref{TruncatedPSF} and \ref{FurtherApprox}, $g(\overline{w},N)$ can be integrated to create the integral probability distributions as a function of $\overline{w}$ and $N$. Note that the expected integral probability distribution is often the same for large regions of the sky map --- for Milagro there is only one such distribution --- and can be tabulated for rapid source detection in step \ref{IDstep}.
  \item \label{IDstep} \textbf{Source Identification --- Section \ref{WATSourceID}.} \textit{Repeated for each potential source location, typically every location on the sky map.}  Compare the observed  $w_{obs}$ and $N_{obs}$ values to the expected background distribution for that location on the sky map, and determine the probability that the observation was produced by a background fluctuation (Equation \ref{baseProb}, Section \ref{TruncatedPSF}, Equation \ref{CorrectProbEq}, or Equation \ref{Prob:Eq} depending on which approximations are being made).  The probability of an observation being produced by the background is typically calculated offline (see step \ref{DistStep}) and loaded into lookup tables.  This makes determining the probability that an observed $w_{obs}$ and $N_{obs}$ pair was produced by a background fluctuation very efficient. 

\end{enumerate}
For a detailed example of a real world implementation, please see \citet{MyThesis} for the application of this analysis method to transient detection with the Milagro observatory.

 The analysis described in this paper uses the available event-by-event  information to achieve a sensitivity similar to that of a well-implemented maximum likelihood analysis while remaining computationally efficient. Additionally, the effects of a number of  different approximations on the sensitivity and speed of the final analysis can be easily determined and tuned to the particular application. 

The analysis technique developed here was designed to fit a particular  niche where the inclusion of event-by-event information can significantly enhance the sensitivity of an analysis and computational  efficiency is an important constraint.  This analysis method is particularly well suited to transient detection in wide field-of-view  gamma-ray observatories, and is currently used for the 40 s -- 3 hour  transient search in the Milagro observatory.

\section{Acknowledgments} 

We are grateful to our many collaborators on the Milagro experiment, past and present, who have helped bring the project to fruition and with whom we have had the pleasure to work.  We also thank the anonymous referee whose thoughtful comments substantially improved the final version of this paper.  MFM was a NASA Graduate  Student Researcher.  This work was also supported by National Science  Foundation Grant PHY-0070927.

\bibliographystyle{elsart-harv} 
\bibliography{WATPaper}

\begin{thebibliography}{8}
\expandafter\ifx\csname natexlab\endcsname\relax\def\natexlab#1{#1}\fi
\expandafter\ifx\csname url\endcsname\relax
  \def\url#1{\texttt{#1}}\fi
\expandafter\ifx\csname urlprefix\endcsname\relax\def\urlprefix{URL }\fi

\bibitem[{Alexandreas et~al.(1993)}]{CygnusTeq}
Alexandreas, D., et~al., 1993. Point source search techniques in ultra high
  energy gamma ray astronomy. Nuclear Instruments and Methods in Physics
  Research A328, 570--577.

\bibitem[{Cassiday et~al.(1989)}]{FlysEyeTeq}
Cassiday, G.~L., et~al., 1989. Evidence for $10^{18}$ ev neutral particles from
  the direction of cygnus x-3. Physics Review Letters 62~(4), 383--386.

\bibitem[{Cassiday et~al.(1990)}]{FlysEyeTeq2}
Cassiday, G.~L., et~al., 1990. Mapping the u.h.e. sky in search of point
  sources. Nuclear Physics B (Proc. Suppl.) 14A, 291--298.

\bibitem[{Feldman and Cousins(1998)}]{FeldmanCousins}
Feldman, G.~J., Cousins, R.~D., 1998. Unified approach to the classical
  statistical analysis of small signals. Physics Review D 57, 3873--3889.

\bibitem[{Hobson and McLachlan(2003)}]{BeyesianRef}
Hobson, M.~P., McLachlan, C., 2003. A bayesian approach to discrete object
  detection in astronomical data sets. Monthly Notice of the Royal Astronomical
  Society 338~(3), 765.

\bibitem[{James(1998)}]{MINUIT}
James, F., 1998. CERN Program Library http://wwwinfo.cern.ch/asdoc/minuit
  /minmain.html.

\bibitem[{Morales(2002)}]{MyThesis}
Morales, M.~F., 2002. A search for tev gamma-ray burst emission with the
  milagrito observatory. Ph.D. thesis, University of California Santa Cruz.

\bibitem[{Woodhams(1989)}]{Woodhams}
Woodhams, M.~D., 1989. Master's thesis: Scintillator data analysis for the
  janzos array. Master's thesis, University of Auckland.

\end{thebibliography}

\end{document}